\documentclass[letterpaper]{article}

\usepackage{xr}
\usepackage[T1]{fontenc}

\usepackage{geometry}
\usepackage{setspace}
\usepackage[articletitle=true]{achemso}

\usepackage{graphicx}
\usepackage{float}
\newfloat{scheme}{htbp}{los}
\floatname{scheme}{Scheme}
\floatname{chart}{Chart}
\newfloat{graph}{htbp}{loh}

\usepackage{chemformula} 
\usepackage[version = 4]{mhchem} 

\usepackage{authblk}
\author[1]{Julien Brodeur}
\author[1]{Laure Sène}
\author[1]{Stéphane Kéna-Cohen*}
\affil[1]{Department of Engineering Physics, Polytechnique Montréal, Montréal}

\title{Mid-infrared distributed-feedback lasing from black phosphorus under nanosecond excitation}
\date{*Email: s.kena-cohen@polymtl.ca}

\begin{document}

\maketitle

\begin{abstract}
Mid-infrared (MIR) light sources compatible with silicon photonics are highly desirable for environmental sensing and free-space optical communications. Conventional GaSb-based quantum-well and interband-cascade lasers, however, rely on complex epitaxial heterostructures, complicating their integration with silicon photonic platforms. Here, we demonstrate a room-temperature MIR surface-emitting distributed-feedback (DFB) laser using black phosphorus (b-P) as the active gain medium. By deterministically integrating exfoliated b-P flakes onto lithographically patterned SiO$_2$ gratings, we obtain narrowband emission with a full width at half maximum below 3 nm under 1064 nm optical excitation. The lasing wavelength is tunable from 3.79 to 4.05$~\mu\text{m}$ through the b-P flake thickness, covering a technologically important region of the mid-infrared. Room-temperature thresholds as low as $(0.25 \pm 0.07)\ \text{mJ/cm}^2$ are achieved under nanosecond excitation. Upon cooling to 110 K, the threshold decreases tenfold to $(0.015 \pm 0.005)\ \text{mJ/cm}^2$.  These results establish planar b-P DFB cavities as a promising platform for heterogeneously integrated MIR lasers based on van der Waals semiconductors.
\end{abstract}

\section*{Keywords}

Black phosphorus, distributed feedback lasers, mid-infrared, two-dimensional materials

\section{Introduction}

Mid-infrared (MIR) optoelectronic devices operating within the $3-5\ \mu\text{m}$ atmospheric window are important for molecular spectroscopy, chemical sensing, medical diagnostics and free-space optical communications \cite{Jung2017,Vodopyanov2020}. Conventional semiconductor sources in this spectral range include GaSb-based type-I quantum well lasers\cite{Shterengas2008,Liang2014}, including multistage cascade designs\cite{Shterengas2014,Hosoda2016}, type-II  InAs/GaInSb interband cascade lasers\cite{Vurgaftman2011,Kim2012} and InP or InAs-based quantum cascade lasers\cite{Liu2010,Lyakh2012}. Although these technologies provide high performance, they rely on compositionally complex epitaxial heterostructures\cite{Tournie2022} that are typically grown on native III–V substrates. Their integration with silicon photonics therefore requires demanding bonding or heteroepitaxial growth processes, motivating alternative gain materials that can be incorporated onto photonic substrates without lattice matching.

Over the past decade, two-dimensional (2D) materials have emerged as compelling active media for on-chip light sources, enabled by their weak out-of-plane van der Waals bonding which allows for heterogeneous integration without lattice-matching constraints\cite{Cheng2021}. Visible and near-infrared lasing has been reported using monolayers of transition metal dichalcogenides coupled to various optical cavities, including microdisks \cite{Ye2015,Salehzadeh2015}, 2D photonic crystals \cite{Wu2015,Li2017,Fang2018,Rong2023}, distributed Bragg reflector microcavities \cite{Shang2017,Koulas-Simos2024}, and metasurfaces \cite{Barth2024}. However, the atomic-scale thickness of monolayer gain media limits their optical-mode overlap and active volume, resulting in low output levels and making conventional lasing signatures difficult to correctly resolve in nanocavity devices.\cite{Lopez-Carrasco2025} Thicker layered semiconductors can retain the heterogeneous-integration advantages of van der Waals materials while providing a substantially larger gain volume.

 Black phosphorus (b-P) is particularly attractive for MIR emission due to its thickness-dependent direct band gap, with a bulk value near 0.3~eV, high carrier mobility, and high photoluminescence quantum yield\cite{Zong2021}. Its strongly anisotropic in-plane optical response is complemented by comparatively weak Auger and surface-recombination losses for a narrow-bandgap semiconductor\cite{Higashitarumizu2023_2}. Accordingly, b-P light-emitting diodes have achieved external quantum efficiencies of several percent\cite{Gupta2022,Brodeur2025}, establishing the material as a promising active medium for MIR light sources.
 
 Optical gain has also been demonstrated in b-P. Huang et al. observed lasing from a lamellar b-P film in an open dielectric cavity\cite{Huang2019}, while Zhang et al. subsequently reported lasing from a b-P flake sandwiched between distributed Bragg reflectors (DBRs)\cite{Zhang2020}. Related vertical-cavity devices based on black arsenic phosphorus extended room-temperature emission from $3.42$ to $4.65\ \mu\text{m}$ through alloy-composition engineering\cite{Zhang2023}. These studies established b-P-based semiconductors as MIR gain media, but operation was demonstrated exclusively under femtosecond excitation, producing highly transient carrier populations. Establishing gain under longer excitation pulses and at reduced carrier densities is important for assessing the prospects for sustained and ultimately electrical injection.

Here, we demonstrate room-temperature MIR b-P surface-emitting distributed-feedback (DFB) lasers operated under 1 ns optical excitation. Exfoliated b-P flakes were deterministically integrated onto lithographically patterned second-order SiO$_2$ gratings, providing in-plane Bragg feedback and surface outcoupling while exploiting the strongly anisotropic optical gain of b-P over an extended propagation length. By varying the b-P flake thickness, we tune the effective index of the fundamental TE$_0$ guided mode and shift the lasing wavelength from 3.79 to 4.05 $\mu$m. A minimum room-temperature threshold fluence of (0.25 $\pm$ 0.07) mJ/cm$^2$ is achieved, corresponding to a threshold carrier density approximately 30 times lower than that inferred from the previous femtosecond-pumped b-P-based report\cite{Zhang2020}. Upon cooling to 110 K, the threshold decreases by an order of magnitude to (0.015 $\pm$ 0.005) mJ/cm$^2$. Together, these results extend 2D-material lasers into the technologically important 3--5~$\mu$m spectral range and highlight b-P as a promising gain medium for heterogeneously integrated MIR light sources.

\section{Results and discussion}

\begin{figure}[htp]
	\centering
	\includegraphics[width=10cm]{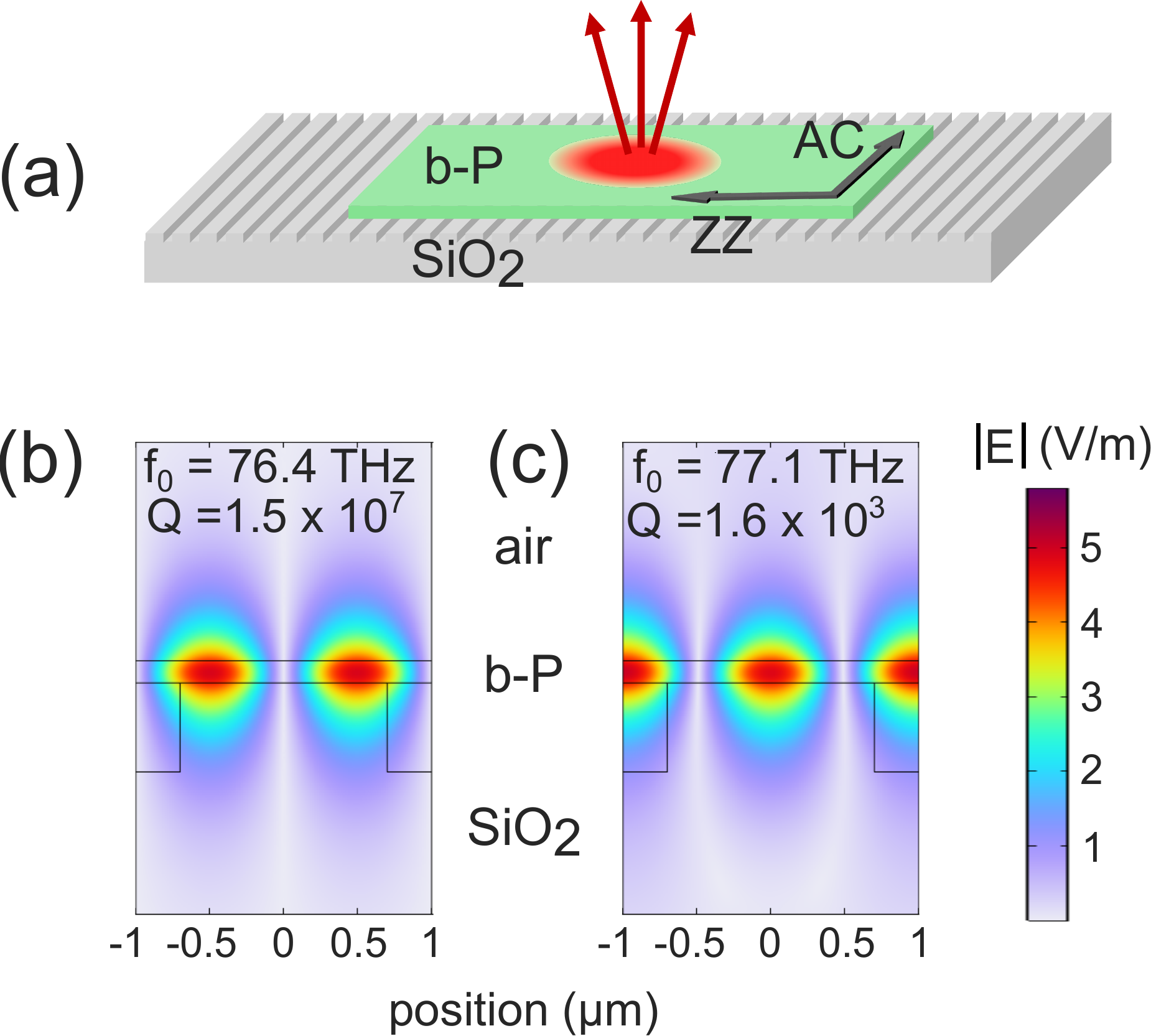}
	\caption{(a) b-P DFB laser structure. Simulated electric field magnitude of the (b) lower branch (dark) and (c) upper branch (radiative) eigenmodes of the b-P DFB structure with periodic boundary conditions at the band edge $k=K_0$. } 
	\label{fig_1}
\end{figure}

The b-P DFB laser architecture is shown in Figure \ref{fig_1}(a). A second-order SiO$_2$/air grating with period $\Lambda=2$~$\mu$m provides in-plane Bragg feedback while coupling the guided mode to surface radiation. The grating was etched 600 nm deep into a 5 $\mu\text{m}$ thermal $\text{SiO}_2$-on-Si substrate with a duty cycle ($D = a/\Lambda$, where $a$ is the ridge width) of 0.7, chosen as a compromise between feedback and radiation coupling \cite{Kazarinov1985}. Following grating fabrication, b-P flakes were isolated via mechanical exfoliation and positioned across the grating using deterministic dry-transfer \cite{Castellanos-gomez2014}. The in-plane armchair (AC) crystallographic axis of the b-P flake was aligned parallel to the grating lines to maximize coupling of the transition dipole moment  with the fundamental transverse electric ($\text{TE}_0$) waveguide mode. Leveraging the intrinsic tendency of b-P to cleave along its structurally robust zigzag (ZZ) axis during exfoliation \cite{Jia2021}, the rectangular flake geometry allowed for its precise spatial orientation by aligning the long ZZ edge perpendicular to the grating lines, a configuration subsequently verified via polarized photoluminescence (PL) spectroscopy. Finally, the device was encapsulated by a 10 nm Al$_2$O$_3$ film grown by atomic layer deposition. 

\begin{figure}[htp]
	\centering
	\includegraphics[width=14cm]{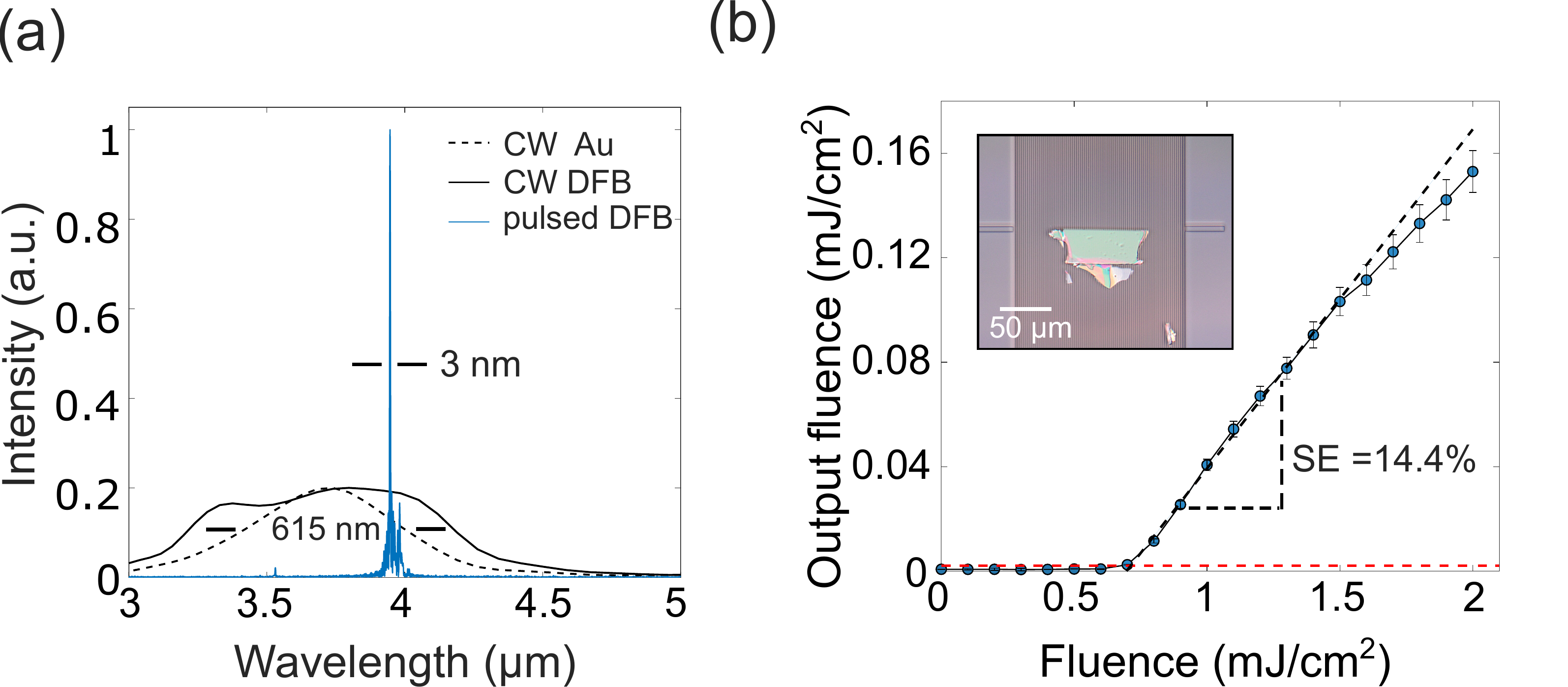}
	\caption{(a) Room-temperature b-P DFB lasing spectrum of the 150 nm thick device under pulsed excitation at an incident fluence 1 mJ/cm$^2$ and CW PL spectra of the same device and of a b-P flake of similar thickness on a gold substrate at a 550 W/cm$^2$ irradiance. (b) Collected output fluence as a function of the pump fluence. The linear section of the curve is used to extract the threshold fluence. The dashed red line indicates the measurement setup noise floor. Inset: optical microscope image of the b-P DFB laser.}
	\label{fig_2}
\end{figure}

An optical micrograph of the fully assembled structure is shown in the inset of Figure \ref{fig_2}(b). To satisfy the second-order Bragg condition for vertical surface emission, the cavity geometry and flake thickness must satisfy the relation \cite{carroll1998distributed}:
\begin{equation}
	\Lambda = \frac{\lambda_B}{n_{eff}}
\end{equation}
where $\lambda_B$ is the target Bragg wavelength and $n_{eff}$ is the effective index of the fundamental TE mode of the air/b-P/\text{SiO}$_2$ waveguide system. To accurately estimate $n_{eff}$, the AC refractive index of the b-P flake was extracted experimentally through polarized white-light reflectivity (see section \ref{sec:SI_index} of the Supporting Information). To target the bulk material gain near $\lambda_B = 3.9\ \mu\text{m}$, a 150 nm flake thickness can be chosen, yielding $n_{eff} \approx 2$, fulfilling the phase-matching condition for a 2 $\mu\text{m}$ grating period. The waveguide effective index as a function of the b-P layer thickness is shown in section \ref{sec:SI_neff} of the Supporting Information.

The two calculated band-edge eigenmodes of an infinite $150\ \text{nm}$ b-P DFB structure at the phase-matching condition $k = K_0 = 2\pi/\Lambda$, are shown in Figures \ref{fig_1}(b) and \ref{fig_1}(c). The complete photonic band structure and corresponding quality ($Q$) factors are provided in Section \ref{SI_fig_eigen} of the Supporting Information. The lower-frequency branch eigenmode (Figure \ref{fig_1}(b)) corresponds to the bound state in the continuum mode. It exhibits even spatial symmetry and is strictly non-radiative, thus yielding an infinite $Q$-factor in the ideal limit. Numerically, we obtain $Q\sim10^7$, limited by the finite precision of the calculation. Conversely, the higher-frequency branch eigenmode (Figure \ref{fig_1}(c)) has odd spatial symmetry and is inherently radiative, giving a finite $Q\sim 10^3$. Because material absorption was omitted from this calculation, losses are entirely governed by surface radiation from the grating. 

To characterize the device behavior below threshold, we measured the room-temperature photoluminescence (PL) spectrum of a 150 nm-thick device under $808\ \text{nm}$ continuous wave at an irradiance of $550\ \text{W/cm}^2$ (Figure \ref{fig_2}(a)). The spectrum is plotted alongside a reference PL spectrum from a b-P flake of similar thickness on a gold substrate. The DFB device retains the broad emission profile of the reference, characterized by a peak near $3.7\ \mu\text{m}$ and a spectral bandwidth of $\sim$ 600 nm, but shows two distinct shoulders at $3.3\ \mu\text{m}$ and $4.1\ \mu\text{m}$ consistent with a grating-induced spectral redistribution of the spontaneous emission. 

Under 1064 nm excitation with 1 ns pulses at 4.5 kHz, the broad spontaneous-emission background gives way to a narrow room-temperature peak above threshold. Figure \ref{fig_2}(a) shows the spectrum acquired at an incident pump fluence of $1\ \text{mJ/cm}^2$. The high-resolution FTIR spectrum exhibits a narrow emission peak with a full width at half maximum (FWHM) < $3\ \text{nm}$ centered at $3.95\ \mu\text{m}$. The emergence of a narrow spectral peak together with a threshold-like input–output response is consistent with lasing. The ringing behavior observed around the primary peak is an artifact of the boxcar apodization used prior to the Fast Fourier Transform (FFT) to maximize the spectral resolution. Note that the sub-threshold behaviour of the device could not be captured under pulsed laser excitation because  the average power emitted by the device remained under the detection threshold of our measurement setup under these conditions. Pumping of a similar b-P flake (170 nm) on an unpatterned SiO$2$ substrate did not yield any luminescence signal even at 6 times the pump fluence used in Figure \ref{fig_2}(a).

 Figure \ref{fig_2}(b) shows the collected laser output fluence as a function of the incident pump fluence. The detector noise floor is demarcated by the red dotted line. A distinct, threshold-like nonlinearity emerges at a critical fluence of $\sim 0.7\ \text{mJ/cm}^2$, beyond which the output power scales linearly with a differential slope efficiency (SE) of (14.4 $\pm$ 0.2) \% before undergoing roll-off near $1.5\ \text{mJ/cm}^2$. The laser achieves a peak output fluence of $\sim$ 0.16 mJ/cm$^2$ corresponding to a peak output power density of 160 kW/cm$^2$ assuming an emission duration comparable to the excitation pulse width.
 
\begin{figure}[htp]
	\centering
	\includegraphics[width=14cm]{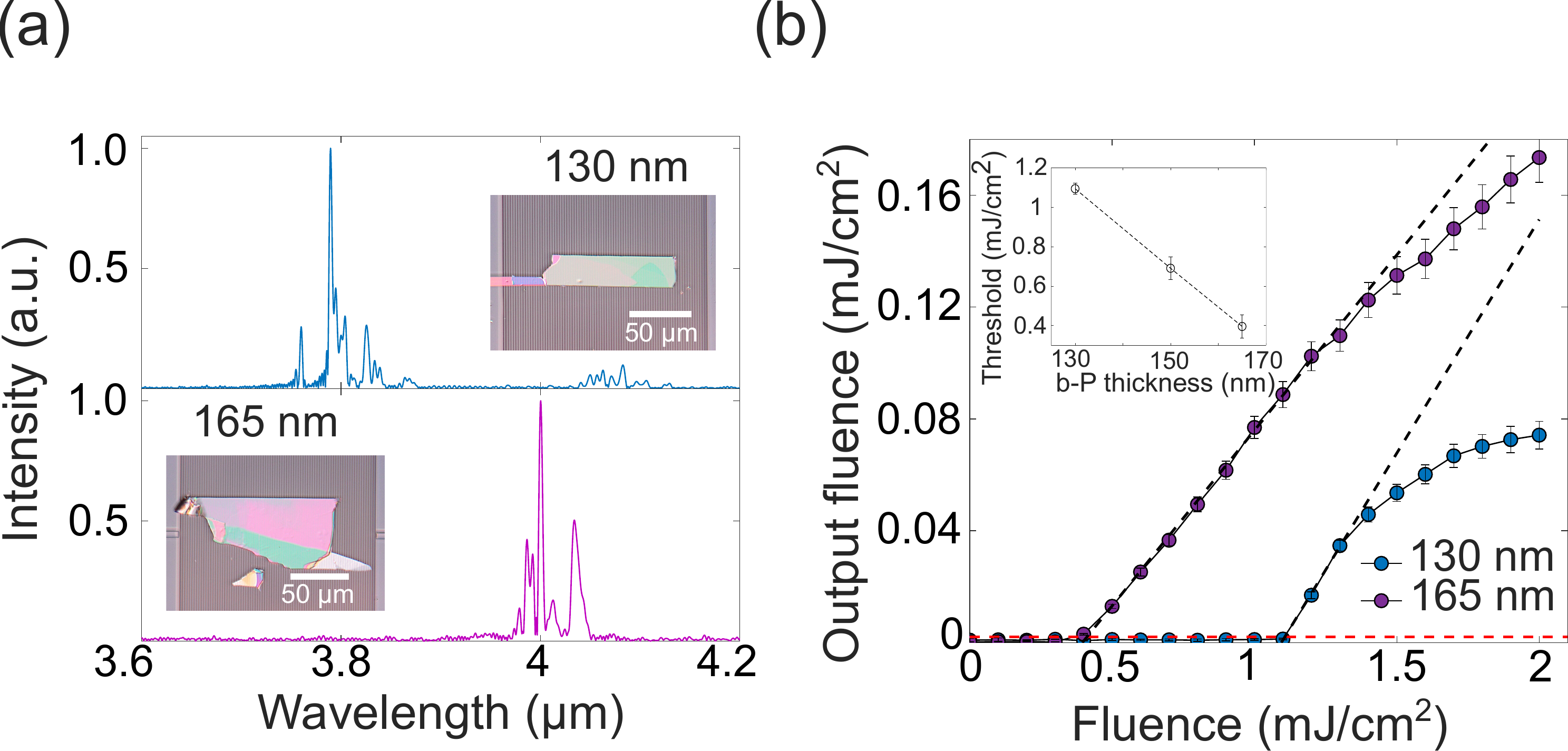}
	\caption{(a) Room-temperature lasing spectra of the 130 and 165 nm-thick b-P DFB lasers at incident pump fluneces of 1.5 mJ/cm$^2$ and 1 mJ/cm$^2$ respectively. Inset: optical micrographs of the b-P DFB lasers. (b) Collected output fluence as a function of the incident pump fluence for the same devices. The linear sections of both curves are used to extract the threshold fluences. Inset: extracted threshold fluence as a function of the b-P flake thickness. }\label{fig_3}
\end{figure}

 To investigate the effect of the b-P flake thickness on the lasing characteristics, we fabricated two additional b-P DFB devices with flake thicknesses of $130\ \text{nm}$ and $165\ \text{nm}$. The high-resolution ($1\ \text{cm}^{-1}$) emission spectra acquired at pump fluences of $1.5\ \text{mJ/cm}^2$ and $1.0\ \text{mJ/cm}^2$ are shown in Figure \ref{fig_3}(a), revealing central lasing peaks at $3.79\ \mu\text{m}$ ($\text{FWHM} = 3\ \text{nm}$) and $4.00\ \mu\text{m}$ ($\text{FWHM} = 5\ \text{nm}$) for the $130\ \text{nm}$ and $165\ \text{nm}$ devices, respectively. Notably, in both spectra, the central DFB mode is accompanied by side peaks. These secondary features are due to longitudinal Fabry-Pérot resonances supported by the finite cavity length formed by the exfoliated b-P flakes. A similar multi-mode behavior emerges in the $150\ \text{nm}$ device when driven at higher pump fluences (see Figure \ref{fig_SI_high_fluence} of the Supporting Information). The origin of these side modes was confirmed by measuring the spectrum of a 170 nm b-P DFB laser with the b-P flake patterned to have parallel edges. In this case, the free spectral range can be clearly attributed to the corresponding longitudinal modes as will be shown below.
 
 The power dependence of the 130 and 165 nm devices is shown in Figure \ref{fig_3}(b). They exhibit a similar linear increase in light emission with slope efficiencies of (18.6 $\pm$ 0.2) \% and (13.4 $\pm$ 0.1) \% respectively following a threshold-like behavior occurring at specific incident fluences. Below these fluences, the emitted power remains below the detection noise floor. As shown in the inset of Figure \ref{fig_3}(b), the extracted threshold values decrease monotonically as a function of b-P flake thickness, reaching a minimum threshold of (0.40 $\pm$ 0.06) mJ/cm$^2$ for the $165\ \text{nm}$ device. These threshold fluences were determined by extrapolation from the fits of the linear regime of the output power curves (fitting coefficients are detailed in Section \ref{sec:SI_lin} of the Supporting Information). 

\begin{figure}[htp]
	\centering
	\includegraphics[width=14cm]{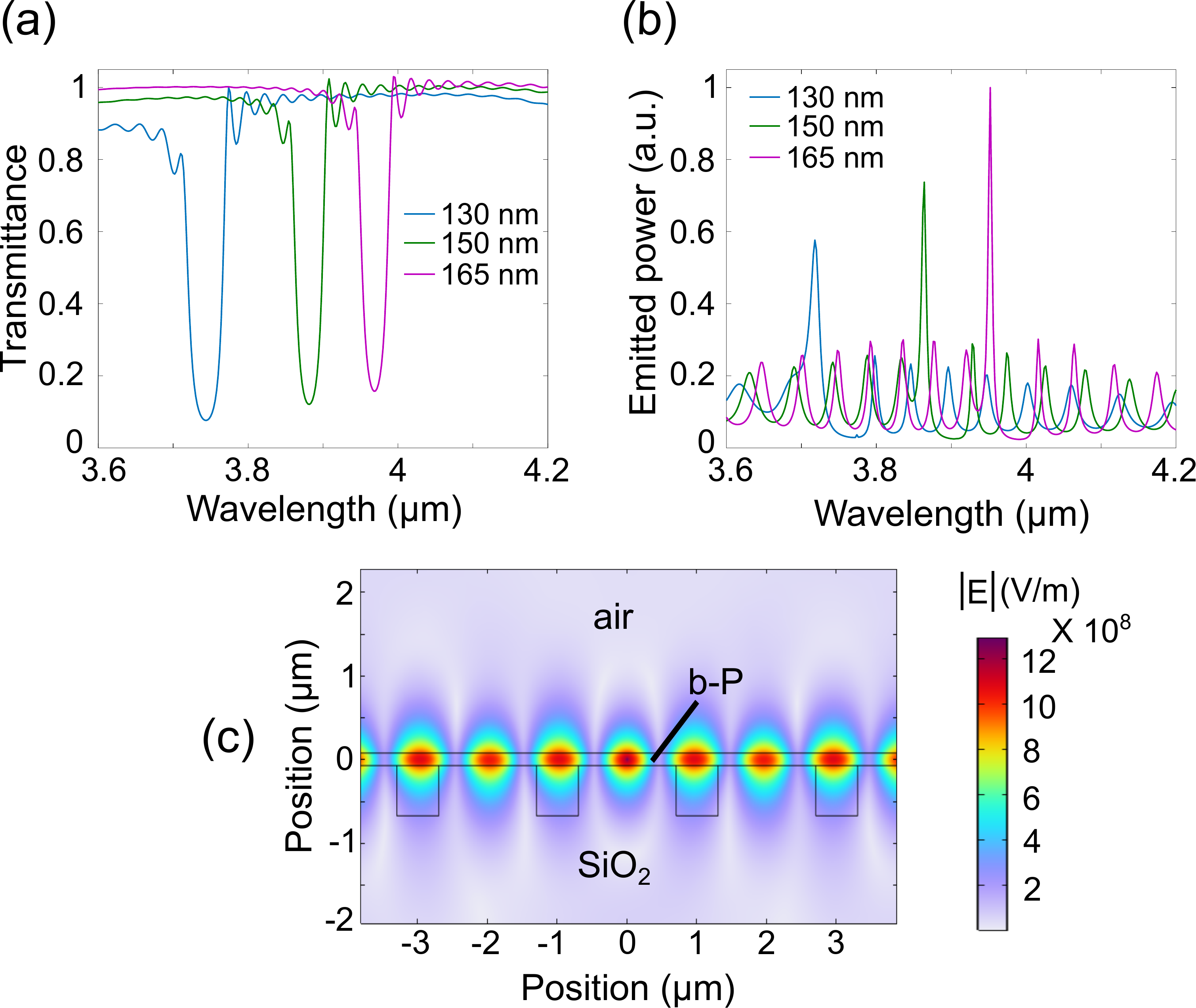}
	\caption{(a) Simulated transmittance spectra of the fundamental TE guided mode for the b-P flake thickness of 130, 150 and 165 nm. (b) Calculated surface emission spectra of an electric dipole oriented parallel to the b-P AC direction for the same flake thicknesses. (c) Simulated electric-field profile of the 150 nm b-P structure at the emission maximum near 3.86 $\mu$m.}
	
	\label{fig_4}
\end{figure}

To understand the origin of our experimental observations, we performed two-dimensional finite element method simulations of the fundamental $\text{TE}_0$ mode transmission through a finite 80 $\mu$m long b-P waveguide/grating architecture as a function of flake thickness. As shown in Figure \ref{fig_4}(a), the center of the photonic stopband redshifts with increasing b-P thickness, which is driven by the enhanced effective index ($n_{\text{eff}}$) of the fundamental guided mode. This simulation redshift follows the experimental trend in central lasing wavelengths. The calculations also reveal a narrowing of the stopband width as the b-P thickness increases. This behavior stems from the stronger confinement of the mode within thicker flakes, which reduces   interaction with the Bragg grating and consequently lowers the second-order coupling coefficient.

To identify the resonances that predominantly couple to the gain medium in the finite structure, we calculated radiation emitted from an electric dipole oscillating parallel to the b-P AC direction, positioned at the geometric center of the structure. Optical pumping was phenomenologically included by introducing a net optical gain coefficient $g = 500\ \text{cm}^{-1}$ (modeled via an imaginary refractive index component, $k = g\lambda / 4\pi$) across the central $40\ \mu\text{m}$ of the b-P layer, matching the experimental pump spot size. Figure \ref{fig_4}(b) shows the radiated power across the upper air boundary as a function of wavelength. The radiated intensity and local field in b-P increase significantly with b-P layer thickness, which directly correlates with the observed reduction in lasing threshold.

Additionally, the spectra exhibit a series of lower-intensity longitudinal Fabry-Pérot modes originating from reflections at the abrupt b-P/air facet interfaces. This result strongly supports our interpretation of the experimental side peaks in Figure \ref{fig_3}(a). Moreover, the maxima from the simulated output spectra occur precisely at the high-frequency edge of the photonic stopband. Figure~\ref{fig_4}(c) shows the simulated electric field norm for the $150\ \text{nm}$ thick device at its peak emission wavelength of $3.86\ \mu\text{m}$. The resulting field distribution closely matches the spatial profile of the odd, upper-branch radiative eigenmode of the infinite structure (Figure \ref{fig_1}(c)) suggesting that the experimentally observed emission preferentially lases via this mode. 

\begin{figure}[htp]\centering\includegraphics[width=14cm]{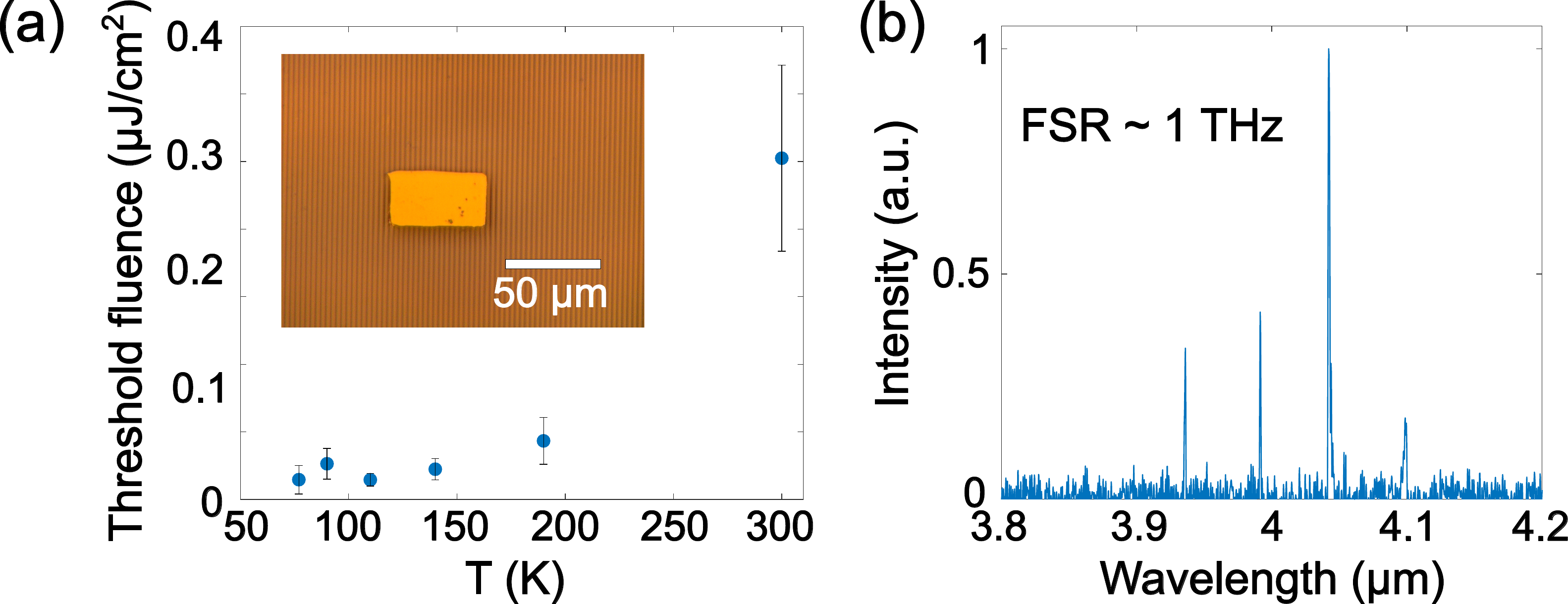}
\caption{(a) Threshold pump fluence of the patterned 170 nm b-P DFB laser as a function of temperature. Inset: optical micrograph of the patterned b-P DFB. (b) Room temperature spectrum of the patterned b-P DFB laser at a 0.5 mJ/cm$^2$ fluence, showing longitudinal Fabry-Pérot modes with a free spectral range (FSR) of approximately 1 THz.}\label{fig_5}
\end{figure}

To investigate the temperature dependence of the lasing threshold, we fabricated an additional b-P DFB device using a 170 nm-thick flake patterned with parallel facets using photolithography and reactive-ion etching prior to transfer onto the grating. This device is therefore distinct from the unpatterned flakes used in the thickness-dependent study of Figure~\ref{fig_3}. The threshold fluences extrapolated from the power dependence curves measured at different temperatures (see Supporting Information Figure \ref{fig:SI_temp}(a)) are shown in Figure \ref{fig_5}(a). At room temperature, the device exhibits a threshold of only (0.25 $\pm$ 0.07) mJ/cm$^2$. A spectrum measured just above threshold at 0.4 mJ/cm$^2$ shows a single lasing peak centered at 4.05 $\mu$m with a FWHM <3 nm (see Supporting Information Figure \ref{fig:SI_temp}b). At a slightly higher pump fluence of 0.5 mJ/cm$^2$ shown in Figure \ref{fig_5}(b), additional longitudinal Fabry–Pérot modes become clearly resolved with a well-defined free spectral range of approximately 1 THz, consistent with the expected mode spacing for the $50~\mu$m-long b-P cavity (see Supporting Information section \ref{SI_sec:FSR} for details).

The room-temperature threshold of the patterned device is lower than that of the unpatterned 165 nm device and does not follow the thickness-dependent trend observed in Figure~\ref{fig_3}(b). This may reflect reduced edge-scattering losses associated with the parallel patterned facets, although device-to-device variability may also contribute. Upon cooling, the threshold decreases by an order of magnitude to (0.015 $\pm$ 0.005) mJ/cm$^2$, as shown in Figure~\ref{fig_5}(a). This decrease begins to saturate below 140K. The initial reduction in threshold is consistent with the increase in peak gain expected when lowering temperature at a fixed carrier density. That increase in gain, however, can compete with a reduction in pumping efficiency due to a reduction in carrier lifetime at lower temperatures. This combination of effects has been shown to give rise to threshold saturation in III-V semiconductor gain media\cite{Hader2008}. 

. 

Because recombination is significant during the 1 ns excitation pulse, we estimate the room-temperature threshold carrier concentration from the quasi-steady-state balance between photocarrier generation and recombination. Assuming bimolecular recombination dominates at threshold and using the parameters described in Supporting Information Section \ref{SI:carrier}, we obtain $n_{th}\sim 1.1 \times 10^{19}\text{ cm}^{-3}$. This value is approximately 30 times lower than the threshold carrier density estimated from Ref.~\citenum{Zhang2020} although differences in pulse duration and carrier dynamics limit a direct comparison. 

The reduced optical threshold is encouraging for eventual electrical injection, although further reductions are required because b-P devices undergo strong efficiency roll-off at high current densities \cite{Higashitarumizu2023,Brodeur2025}. Given our lowest room-temperature threshold fluence of 0.25 mJ/cm$^2$, we can estimate the threshold current density required for lasing to be $\sim$ 100 kA/cm$^2$ which is $\sim$20 times higher than the current densities before which severe roll-off occurs\cite{Brodeur2025}. In contrast, the 0.015 mJ/cm$^2$ threshold measured at 110 K would correspond to an equivalent current density of only $\sim$ 6 kA/cm$^2$ if the same generation–recombination conversion were assumed. Although temperature-dependent recombination and injection processes preclude a direct quantitative comparison, this value suggests that cryogenic electrical pumping may already be within reach.

At room temperature, we can estimate a b-P transparency carrier concentration of $n_{tr} = 3 \times 10^{18}\text{cm}^{-3}$ using the Bernard-Durrafourg condition with an effective density of states model (details are in Supporting information section \ref{SI:carrier}), which is 4 times lower than the estimated threshold carrier density. In the high injection regime where bimolecular recombination dominates, and assuming a unity injection efficiency, the threshold current density can be approximated by

\begin{equation} \label{eq:Jth}
	J_{th} \simeq q t_{act} B \left(n_{tr} + \frac{\alpha}{\Gamma\gamma} \right)^2 
\end{equation} 

where $q$ is the elementary charge, $t_{act}$ is the active layer thickness, $B$ is the bimolecular recombination coefficient, $\alpha$ is the resonator loss coefficient and $\gamma$ is the gain coefficient where the material gain is given by $g(n)=\gamma(n-n_{tr})$ and $\Gamma$ is the modal confinement factor. In the regime where the second term dominates, the threshold current density scales quadratically with the resonator losses meaning that a significant reduction of $J_{th}$ is possible by either improving the optical quality of the grating to reduce scattering losses or using a transparent grating material such as sapphire to reduce absorption losses (SiO$_2$ is slightly absorptive at 4 $\mu$m). In the low-loss limit, where $n_{th}$ approaches $n_{tr}$, Eq. \ref{eq:Jth} predicts a linear dependence of the threshold current density on active layer thickness $t_{act}$. In this regime, the use of a double heterostructure to confine carriers within a thinner b-P flake can also be used to further reduce $J_{th}$.

In conclusion, we have demonstrated room-temperature MIR lasing from unalloyed black phosphorus under 1 ns optical excitation using a planar surface-emitting DFB cavity. Deterministic integration of exfoliated b-P onto patterned SiO$_2$ gratings provides optical feedback and surface outcoupling without requiring epitaxial growth or vertical dielectric mirror stacks. The lasing wavelength can be varied from $3.79\ \mu\text{m}$ to $4.05\ \mu\text{m}$ through the b-P flake thickness, while increased optical confinement in thicker flakes is accompanied by a reduction in threshold across the devices studied.

A minimum room-temperature threshold fluence of $(0.25 \pm 0.07)\ \text{mJ/cm}^2$ is obtained for a patterned 170 nm-thick device. This corresponds to a threshold carrier density approximately 30 times lower than that inferred for the previous femtosecond-pumped b-P laser\cite{Zhang2020}. Cooling further reduces the threshold by an order of magnitude to ($0.015 \pm 0.005)\ \text{mJ/cm}^2$ at 110 K, approaching an equivalent injection-current regime relevant to cryogenic electrical pumping. These results extend b-P lasing from ultrafast excitation toward substantially longer carrier-injection timescales and identify black phosphorus as a promising gain material for heterogeneously integrated, ultimately electrically driven MIR light sources.
 
\section{Methods}
\subsection{Device fabrication}
The gratings were patterned in photoresist on 5 $\mu$m SiO$_2$/Si substrates using laser lithography (Raith Picomaster 150). The gratings were formed by etching the SiO$_2$ by ICP-DRIE (Plasmaterm CORIAL 210IL) using a CHF$_3$/O$_2$ (50/5 sccm) gas mixture with  1500/100 W ICP/RF power for 2 minutes. The residual resist was removed by cleaning the substrates in a piranha solution for 10 minutes. The b-P flakes were mechanically exfoliated and directly transferred on the gratings using a PDMS stamp. Thickness of the flakes was determined using white light reflectivity and transfer matrix modeling during the fabrication process. To encapsulate the device, a 10 nm Al$_2$O$_3$ layer was deposited using atomic layer deposition (Anric) using trimethylaluminum (TMA) and water as precursors at a 150$^{\circ}$C temperature. All fabrication steps were performed under a nitrogen atmosphere with oxygen and water vapor levels below 0.1 ppm.
\subsection{PL and lasing measurements}
CW PL spectra were obtained by pumping the b-P devices with a 808 nm laser diode at a 550 W/cm$^2$ irradiance. The CW laser diode signal was modulated at 500 Hz using a mechanical chopper.  Lasing spectra were obtained by pumping the devices with a Nd:YAG laser at a 4.5 kHz repetition rate. Both CW and lasing signal were collected with a CaF$_2$ lens with 20 mm focal length and redirected to a FTIR spectrometer operated in step scan mode equipped with a liquid nitrogen cooled InSb detector. The detector output was sent to a lock-in amplifier using either the mechanical chopper frequency or the Nd:YAG repetition rate as the reference signal. The lock-in amplifier output was finally sent to the FTIR analog to digital converter for processing. Acquired spectra and output fluence curves were corrected for absolute detector response. The devices were tested under vacuum conditions to prevent degradation.        
   
\section*{Acknowledgements}
Funding for this work was provided by the Natural Sciences and Engineering Research Council of Canada and Canada Research Chairs Program. We would like to acknowledge CMC Microsystems, manager of the FABrIC project funded by the Government of Canada for the COMSOL licence and  and fabrication support. We would also like to thank (CM)$^2$ for the cross-section SEM images.

\section*{Supporting information}

AC refractive index measurement of b-P flakes, effective index of the air/b-P/SiO$_2$ waveguide, eigenfrequency and eigenmode simulation of b-P DFB with periodic boundary conditions, cross section SEM image of the 150 nm b-P DFB laser, linear fit parameters of the output fluence curves, lasing spectrum of the 150 nm b-P DFB device at high fluence, threshold fluence temperature dependence, estimation of threshold and transparency carrier concentrations.


\bibliography{reference.bib}

\end{document}